\documentstyle[12pt,aasms4]{article}		

\slugcomment{V1.0 for ApJ, 990301}

\begin{document}

\title{The Spite Lithium Plateau: Ultra-Thin but 
Post-Primordial\footnote{Based on
observations obtained with the University College London \'echelle 
spectrograph (UCLES) on the 3.9~m Anglo-Australian Telescope (AAT), the
Double Beam Spectrograph (DBS) on the Australian National University 2.3m
telescope, and the Utrecht \'echelle spectrograph (UES) on the 4.2~m
William Herschel Telescope (WHT).}}

\author{Sean G. Ryan}
\affil{Royal Greenwich Observatory (closed), and 
Institute of Astronomy, Madingley Road, Cambridge CB3 0HA, United Kingdom;
email: sgr@ast.cam.ac.uk}

\author{John E. Norris}
\affil{Research School of Astronomy and Astrophysics, The Australian National 
University, Private Bag, Weston Creek Post Office, ACT 2611, Australia;
email: jen@mso.anu.edu.au} 

\and

\author{Timothy C. Beers}
\affil{Department of Physics and Astronomy, Michigan State University, East 
Lansing, MI 48824;
beers@pa.msu.edu}

\begin{abstract}

We have studied 23 very metal-poor field turnoff stars, specifically chosen to
enable a precise measurement of the dispersion in the lithium abundance of the
Spite Li plateau.  We concentrated on stars having a narrow range of effective
temperature and very low metallicities ([Fe/H]~$^<_\sim$~$-2.5$) to reduce the
effects of systematic errors, and have made particular efforts to minimise
random errors in equivalent width and effective temperature. A typical formal
error for our abundances is 0.033~dex (1$\sigma$), which represents a factor of
two improvement on most previous studies.

One of the 23 stars, G186-26, was known already to be strongly Li depleted.  Of
the remaining 22 objects, 21 (i.e. 91\% of the original sample) have abundances
consistent with an {\it observed} spread about the Spite Li plateau of a mere
0.031~dex (1$\sigma$).  As the formal errors are 0.033~dex, we conclude that
the intrinsic spread $\sigma_{\rm int}$ is effectively zero at the very
metal-poor halo turnoff.  (Inclusion of the twenty-second star would inflate
the observed spread to only 0.037~dex, leaving $\sigma_{\rm int} < 0.02$.)
Furthermore, we have established this at a much higher precision than previous
studies ($\sim$0.06--0.08~dex).

Our sample does not exhibit a trend with effective temperature, though the
temperature range is limited.  However, for $-3.6~<$~[Fe/H]~$<~-2.3$ we do
recover a dependence on metallicity at d$A$(Li)/d[Fe/H]~=~$0.118 \pm
0.023(1\sigma)$~dex per dex, almost the same level as discussed previously.
Earlier claims for a lack of dependence of $A$(Li) on abundance are shown to
have arisen, in all likelihood, from the use of noisier estimates of effective
temperatures and metallicities, which have erased the real trend.  The
dependence is concordant with theoretical predictions of Galactic chemical
evolution (GCE) of Li (even in such metal-poor stars) and with the published
level of $^6$Li in two of the stars of our sample, which we use to infer the
GCE $^7$Li contribution. The essentially zero intrinsic spread ($\sigma_{\rm
int} < 0.02$~dex) inferred for the sample leads to the conclusion that either
these stars have all changed their surface Li abundances {\it very} uniformly,
or else they exhibit close to the primordial abundance sought for its
cosmological significance.  Although we cannot rule out a uniform depletion
mechanism, economy of hypothesis supports the latter interpretation.  The lack
of spread in the $A$(Li) abundances limits permissible depletion by current
rotationally-induced mixing models to $<$ 0.1~dex.

Correcting for the GCE contribution to both $^6$Li and $^7$Li, we infer a
primordial abundance $A$(Li)$_p$~$\simeq$~2.00~dex, with three systematic
uncertainties of up to 0.1~dex each depending on uncertainties in the effective
temperature scale, stellar atmosphere models, and correction for GCE.  (The
effective-temperature zeropoint was set by Magain's and Bell \& Oke's $b-y$
calibrations of metal-poor stars, and the model atmospheres are Bell's, without
convective overshoot.) We predict that observations of Li in extremely
low-metallicity stars, having [Fe/H]~$<$~$-3$, will yield smaller $A$(Li)
values than the bulk of stars in this sample, consistent with a low primordial
abundance.

The difference between our field star observations and the M92 data in the
literature suggests that real field-to-cluster differences in Li evolution may
have occurred. This may indicate different angular momentum evolutionary
histories, with interactions between protostellar disks in the dense globular
cluster environments possibly being responsible.  Further study of Li in
globular clusters and in very metal-poor field samples is required to clarify
the situation.

\end{abstract}


\vfill
\eject

\section{Introduction}					

Beginning with Spite \& Spite (1982), many authors have used the apparent
uniformity of the abundance of lithium in the atmospheres of metal-poor ([Fe/H]
$< -1.0$) subdwarfs warmer than $T_{\rm eff}~=~5600$~K to infer the primordial
value generated by standard Big Bang nucleosynthesis.  By restricting a sample
to $T_{\rm eff} > 5600$~K, one avoids well-documented processes which alter the
surface Li abundance in cooler dwarf stars (e.g.  Deliyannis, Demarque, \&
Kawaler 1990).  If this interpretation is correct, the so-called ``Spite Li
Plateau'' abundance of $A({\rm Li}) = 12 + {\rm lg} (N({\rm Li})/N({\rm H}))
\simeq 2.1$ provides constraints on the baryon-to-photon ratio in the early
universe, and hence $\Omega_b$ (e.g. Deliyannis 1995)\footnote{lg $X$ =
log$_{10}$ $X$}.

However, several theoretical and observational results have cast doubt on use
of the observed Li plateau abundance as the primordial value. Lithium is
fragile, and some stellar evolutionary models show that Li could have been
depleted by an order of magnitude from a high primordial value and still
attained plateau-like abundances by the age of the halo (e.g. Pinsonneault,
Deliyannis, \& Demarque 1992). More recent computations by Pinsonneault et al.
(1998), using an improved treatment of angular momentum evolution and
comparisons with more modern observations, have reduced the permissible $^7$Li
depletion to the range 0.2--0.4~dex.  Trends of lithium abundance with $T_{\rm
eff}$ and [Fe/H] (effectively a tilted plateau) have also been measured
(Thorburn 1994; Norris, Ryan, \& Stringfellow 1994; Ryan et al. 1996a) which
would not exist if the Li were primordial, though these results may be driven
by larger than expected systematic errors in the effective temperatures
(Bonifacio \& Molaro 1997).  The huge ($>$~1~dex) Li deficiencies in some stars
which are otherwise indistinguishable from normal plateau stars (Hobbs, Welty,
\& Thorburn 1991; Thorburn 1994; Norris et al. 1997a; Ryan, Norris, \& Beers
1998) highlight the incompleteness of our understanding of Li processing in
halo stars.

A direct challenge to the thesis of a primordial, and therefore uniform, Li
plateau was mounted by Deliyannis, Pinsonneault, \& Duncan (1993), and
supported by Thorburn (1994), who argued that the spread in measured
plateau-star abundances exceeds that expected from observational errors.
Deliyannis et al. tabulated a range of dispersions, depending on the
characteristics of the sample, but with a  {\it minimum} spread of $\sigma =
0.04$~dex, while Thorburn (1994) found a value around 0.1~dex for a much larger
sample.  Both groups concluded that Li production and/or depletion mechanisms
had operated prior to the birth or during the evolution of the stars, in which
case the measured Li abundance would not reflect solely that from Big Bang
nucleosynthesis.  Large ranges in Li abundance have also been deduced for
subgiants in M92 (Deliyannis, Boesgaard, \& King 1995; Boesgaard et al. 1998).
As further evidence of star-to-star differences in the halo field, Ryan et al.
(1996a) cited the three stars G64-12, G64-37, and CD$-33^\circ1173$, all of
which have extremely low metallicities ([Fe/H]$ < -3$), are apparent
non-binaries, and have surface temperatures $T_{\rm eff} \approx 6250$ K, but
for which they computed abundances $A({\rm Li}) = 2.29 \pm 0.05$, $2.01 \pm
0.04$, and $1.89 \pm 0.06$, respectively.

However, the case for a measurable dispersion in the Li plateau has not gone
unchallenged. Most recently, Molaro, Primas, \& Bonifacio (1995), Spite et al.
(1996), and Bonifacio \& Molaro (1997) have questioned whether some of the
error estimates in earlier works were realistic, and have suggested that the
dispersion is no greater than 0.08--0.10~dex, less than that found by Thorburn
(1994), but not excluding the smaller scatter of Deliyannis et al. (1993).
Ryan et al. (1996a) noted that most, but not all, published measurements could
be reconciled within their claimed errors, thus illustrating that some error
estimates were optimistic, a result which biases one towards over-interpreting
the spread about the mean plateau value.

We set out to provide a substantially more accurate assessment of scatter about
the Li plateau, to see whether we could rule out a purely primordial
interpretation, or whether the width was essentially consistent with small
uncertainties in the measurements and analysis.  We note at the outset that a
very thin plateau is a necessary but not sufficient condition for the observed
abundance to be primordial.  The Li plateau may be of infinitesimal width, but
depend on effective temperature and/or metallicity, in which case it will still
not provide the primordial abundance, though some stars may be very close to
it.

\section{Definition of the Sample}

We sought very high signal-to-noise (S/N) measurements of the 6707~\AA\ Li
doublet in a group of well-selected halo stars, with the aim of measuring Li
abundances to higher precision than had been routinely accomplished previously.

Estimates of the effective temperatures of stars are notoriously uncertain,
particularly absolute as opposed to relative estimates, yet derived Li
abundances depend on temperature.  For example, the existence of
temperature-dependent trends in the Li plateau depends on which effective
temperature scale is adopted (compare Ryan et al. (1996a) and Bonifacio \&
Molaro (1997)).  Uncertainties in star to star abundance comparisons also
increase if their temperatures differ, because stellar atmosphere structures
and color-effective temperature transformations also depend on temperature.  To
minimise the effects of systematic errors, we restricted our sample to a very
narrow range in $T_{\rm eff}$, and chose a narrow metallicity regime
([Fe/H]~$^<_\sim$~$-2.5$), since this avoids possible metallicity-dependent
errors in the color-effective temperature transformation and model stellar
atmospheres.  Our sample targeted effective temperatures in the range $6100{\rm
K}\; (\pm 50) < T_{\rm eff} < 6300{\rm K}\; (\pm 50)$, 
metallicity in the range $-3.5 \le {\rm [Fe/H]} \le -2.5$.  We also restricted
our sample to stars brighter than V~=~13 because of the requirement for high
S/N.

The narrow temperature range places the stars at the turnoff of an old
main-sequence population, effectively eliminating subgiants which spent their
main-sequence lives at higher temperatures than those now observed, and hence
have different evolutionary histories.  Another benefit of this restriction is
that the surface gravity of the sample covers only a narrow range at the
turnoff, though it has been noted many times previously that the Li abundance
derived for halo dwarfs is quite insensitive to surface gravity errors.  The
use of very low metallicity stars means that we are sampling material which has
undergone a minimum of nucleosynthetic processing since the Big Bang.

We developed a target list of approximately 30 stars from the surveys of
Schuster \& Nissen (1988), Ryan (1989), Beers, Preston, \& Shectman (1992), and
Carney et al. (1994).  A large sample was sought to reduce the impact of one or
two ``pathological'' objects, such as marginally depleted examples of the
ultra-Li-depleted stars or unrecognised binaries.  We hoped to make multiple
measurements of each one to verify the repeatability and to provide a check for
radial velocity variability.

With the available telescope time, 22 of the stars were observed, and these are
presented in Table~1. Also included in the table is G186-26, a known
ultra-Li-depleted star (Hobbs, Welty, \& Thorburn 1991) which satisfied our
selection criteria, but which we chose not to reobserve since its surface Li
deficiency is already well established. Its relevance to our work is as a
reminder that at least some otherwise similar stars have depleted the vast
majority of their Li.

\section{Basic Data}

The stars have Johnson-Cousins photometry from a small number of sources
referenced in Table 1, whose consistency and accuracy have already been
established at $\sigma$~=~0.010~mag per observation for B--V and R--I, and
$\sigma$~=~0.007~mag in V--R (Ryan 1989).  Str\"omgren photometry from
Schuster, Nissen, and collaborators (see references in table) is available for
all but one star.  The columns headed n$_J$, n$_S$, and n$_\beta$ in the table
give the number of BVRI, uvby, and $\beta$ measurements respectively.  Multiple
measurements improve the photometric accuracy, which is important in deriving
effective temperatures for the stars.  We reduce the Johnson errors to 0.007
(for B--V and R--I) and 0.005 (for V--R) for two or more observations.  Where
rounding errors of up to 0.005~mag affect Johnson-Cousins colors quoted to only
0.01~mag, we adopt larger uncertainties --- 0.015 (for B--V and R--I) and 0.010
(for V--R) for single measurements, and 0.010 (for B--V and R--I) and 0.007
(for V--R) for two or more measurements.  Schuster \& Nissen (1988) quote mean
errors less than 0.008~mag for $b-y$ and 0.011~mag for $\beta$ where there are
three observations per star.  Given that all of our program stars have 3 or
more Str\"omgren observations, we adopt these error estimates for our entire
sample.

Estimates of the interstellar reddening have been obtained from two techniques.
Values estimated from reddening maps (Lucke 1978; Burstein \& Heiles 1982) and
Johnson photometric distances have been made by Carney et al. (1994) and Ryan
(1989), and are listed in Table~1 as E(BV).  Str\"omgren photometry estimates
of E($b-y$) are based on a comparison of the $b-y$ color and $\beta$
reddening-free index (Schuster \& Nissen 1989, eq. (1)), and these values are
tabulated under E(by).  Based on the central wavelengths of the bandpasses and
a $1/\lambda$ reddening law, a relation E($b-y$) = 0.7$\times$E(B--V) is
expected.  However, comparison of the E(B--V) inferred from the Str\"omgren
values with the map-based values shows that the former are higher by 0.020~mag.
The Str\"omgren technique suggests a mean reddening for the sample of $\langle
E(B-V) \rangle$~=~0.035~mag, whereas the map-based values suggest $\langle
E(B-V) \rangle$~=~0.015~mag.  We lack solid  evidence as to which reddening
scale --- map or Str\"omgren --- is better, but as the sample is fairly bright,
we expect the intrinsic reddening to be low, so reduce all Str\"omgren values
of E(B--V) by 0.020~mag prior to averaging\footnote{We cannot discount the
possibility that the Galaxy does indeed have a high local reddening. This has
been suggested already by Schuster et al. (1996) who find, on the basis of
Str\"omgren photometry, an average reddening of 0.036 within 30$^\circ$ of the
South Galactic Pole.}.  Once this offset is taken into account, the RMS error
inferred for a single E(B--V) estimate is 0.009~mag. This error is assumed to
affect all dereddening vectors of non-zero magnitude.  We adopt
E(V--R)~=~0.78$\times$E(B--V) and E(R--I)~=~0.82$\times$E(B--V) following
Savage \& Mathis (1979).

Table~1 also records measurements of the H$\delta$ line spectroscopic index,
HP2, from Beers et al. (1999) supplemented with new, 1\AA -resolution, high S/N
determinations based on observations with the 2.3m telescope on Siding Spring
Mountain in 1998 March and September. This pseudo-equivalent width index
complements the $\beta$ index and helps establish the effective temperature
scale (below). It has the benefit of being independent of reddening,
essentially independent of metallicity for our metal-poor sample, and having
better temperature sensitivity than $\beta$ for the temperatures of our sample.

\section{Spectroscopic Observations and Data Reduction}

\subsection{Observational Program}

Previous investigations of the Li plateau (see \S1) claimed the significance of
spreads at levels $\sigma$~$\sim$~0.08--0.10~dex, but Deliyannis et al. (1993)
showed that the Li spread could be as small as $\sigma$~=~0.04~dex, depending
on which subsample of stars they analysed.  These values indicated that we
would require accuracies of order $\le$10\%, or $\le$0.04~dex, to clarify the
situation.  The equivalent width for the Li line in halo turnoff stars is
$\sim$~20~m\AA, thus requiring ${\sigma_W}~^<_\sim$~2~m\AA.  This in turn
demanded high resolution spectra (R $\sim$ 40000, which just resolves the
6707~\AA\ Li doublet) and high S/N.

Observations were made using the University College London \'echelle
spectrograph (UCLES) at the coud\'e\ focus of the Anglo-Australian Telescope,
spread over four epochs (in many instances utilising partial nights).  Two
different observers acquired the data as follows: August 28, 1996 (SGR); August
18--23, 1997 (JEN); April 8--10, 1998 (JEN); and August 10--15, 1998 (SGR).
Although cross-dispersed \'echelle spectra often have only limited spatial
coverage which makes the sky and scattered light level difficult to measure, we
used UCLES with its 79~lines~mm$^{-1}$ grating, which gives a 14~arcsec slit
length, thus providing an unambiguous background subtraction.  The spectra are
shown co-added for multiple epochs in Fig.~1.  For the northern star
BD+9$^\circ$2190, we had only one low S/N (=85) measurement with the AAT, so we
obtained a supplementary observation in service time with the Utrecht \'echelle
spectrograph (UES) at the Nasmyth focus of the William Herschel Telescope, on
November 5, 1998. The UES is almost identical to the UCLES.

The Li measurements (discussed in detail in \S4.2) are presented in Table 2,
where we tabulate for each epoch: the S/N, the equivalent width, and the
equivalent width error.  The S/N is taken as the lesser of that expected from
Poisson photon statistics and the scatter actually measured about the continuum
fit.  The variance$^{-1}$-weighted sum, $\bar{W}$, and error
$\sigma_{\bar{W}}$, are also provided, as are two [Fe/H] values and our
estimates of the effective temperature (see \S5).  The first column of [Fe/H]
values is from the literature and derives from high- and/or medium-resolution
spectroscopic observations, for which the errors are believed to be
$\sigma$~$\simeq$~0.15~dex (see references in table).  The second [Fe/H]
entries were obtained by applying the calibration of Beers et al. (1999) to the
1~\AA-resolution spectra from which the HP2 index was measured. The agreement
between the two sets of values is very good; we shall return to this point
later in the discussion.

To obtain the precision needed to examine potentially small levels of scatter
about the Spite Li plateau, it was clear that we would need very accurate
equivalent width measurements.  In an earlier work (Norris~et~al. 1994) we
estimated our Li equivalent width uncertainties due to random noise as
$\sigma(W) = 150/({\rm S/N}_{50})$ m\AA, where the S/N was per 50~m\AA\ pixel.
In the current study we sum over a wider band to be more certain of including
all of the line (see below), and so derive a larger numerator giving $\sigma(W)
= 184/({\rm S/N}_{50})$ m\AA.  As most epochs (except the single-night pilot
run in 1996) have S/N in excess of 100, we expected to achieve accuracies
better than 2.0~m\AA\ per observation.

High absolute accuracy is harder to achieve that high internal precision.  We
discuss internal and external errors below, emphasising that our primary
requirement in studying scatter about the Li plateau is a large, homogeneously
selected, consistently reduced, precisely measured, and consistently analysed
set of data.  Several procedures were adopted to identify and minimise errors
in order to meet this requirement.  Firstly, we sought two epochs of data on
each star to permit us to verify the repeatability of each measurement; we
obtained multiple observations for 12 of the 22 stars.  Secondly, all raw data
were reduced by two of us independently, using different software. This allowed
us to verify that the reduced spectra were consistent irrespective of which
software, algorithms, and personal techniques were applied.  Finally, two
different techniques were used to measure the equivalent widths from the
reduced spectra.

\subsection{Details on Equivalent Width Measurements}

\subsubsection{Continuum Placement}

The spectra of very metal-poor main-sequence-turnoff stars are essentially
devoid of lines over the range 6700--6715~\AA\ apart from the 6707~\AA\ Li
doublet itself.  Even the \ion{Ca}{1} line at 6717~\AA\ is invisible at the low
metallicity and warm temperatures of many of these objects.  (An equivalent
width $W_{6717} < 1$~m\AA\ is expected for a model with $T_{\rm eff} = 6000$~K,
lg~$g$~=~4.0, and [Fe/H]~=~$-3$.) The continuum can therefore be defined
accurately and objectively by fitting the mean flux on either side of the Li
doublet.  Two techniques for measuring equivalent widths are described in the
following subsection. For the direct summation method, the continuum was
computed as a quadratic fit to the flux in zones 2.5~\AA\ wide on either side
of a 1.2~\AA -wide zone of avoidance centred on the Li feature.  The
Gaussian-fitting technique employed a linear continuum, again using the mean
flux on either side of the Li feature, though the exact width of each continuum
zone (approx. 4~\AA) was allowed to vary from star to star.

\subsubsection{Equivalent Width Calculations}

The Li~6707~\AA\ doublet separation is quite large (0.15~\AA), so for the
narrow range of line strengths in our program stars, the FWHM of the spectral
feature is not very sensitive to the instrumental resolution, which was in any
event constant throughout the observing program.  We also confirmed that the
line broadening of the Mg 'b' lines was similar in all objects, as a check
against rotational broadening or the presence of a barely-resolved spectrum of
a secondary companion.  As a result of the similarity of our program stars, the
actual width of the doublet line is expected to be constant for all of them,
with only the line depth responding to equivalent width differences.

Once the continuum was defined, equivalent width measurements were made in two
ways.

The first measurement technique was to centroid on the Li doublet, and then
compute the equivalent width from the residual flux summed within a band $\pm
0.34$~\AA\ of that centroid. The width of this band was set considering the
known width of the doublet and resolving power of the spectrograph, and
confirming on the spectra that this was a sensible choice.  The very high S/N
observations of HD~140283, whose Li equivalent width (48~m\AA) is larger than
the program stars and thus sets an upper limit on the FWHM of the doublet in
our (hotter) stars, showed that $<$1\% of the absorbed flux would be missed
over a band width of $\pm$0.34~\AA. At the same time we avoid unwanted
sensitivity to noise fluctuations that would arise if we summed over more
pixels than necessary.

The second technique involved performing a Gaussian fit, but with the Gaussian
FWHM fixed at 0.305~\AA, again determined from the HD~140283 observations.  As
noted above, since the Li line is weak and the doublet resolved, its FWHM is
determined by the doublet separation and the instrumental profile rather than
the equivalent width.  This procedure was adopted to avoid having noise in the
line cause unphysical line widths in the fit.

Measurement of a given spectrum with the two equivalent width techniques
(direct summation and constrained Gaussian fitting) showed good agreement.  For
the 1997 data, the mean difference between measurements and its standard
deviation was $\langle W_{\rm sum} - W_{\rm Gauss}\rangle$ = $-0.3$~m\AA, with
$\sigma$~=~1.7~m\AA.  Similarly, for the 1996 data, the mean difference was
$\langle W_{\rm sum} - W_{\rm Gauss}\rangle$ = $-0.1$~m\AA, with
$\sigma$~=~1.8~m\AA.  This gives us confidence that the two techniques
introduce no significant systematic differences. (This test was not repeated in
1998, as there had been no changes to the procedures.)

As noted above, two authors reduced the data independently.  Once we were
satisfied that both equivalent width measurement techniques gave consistent
results, one approach was applied by one author to his spectral reductions, and
the second technique was applied by the other. The average of the two
measurements was then adopted for each epoch.

\subsubsection{Internal Errors}

Our error estimates are based on the random noise accumulated over the width of
the line (e.g. Cayrel 1988).  For our pixel spacing and the width over which we
measure the line in the direct summation method, we obtain the relationship
$\sigma(W) = 184/({\rm S/N}_{50})$ m\AA\ where S/N$_{50}$ is for a 50~m\AA\
pixel.  Before utilising this model in the abundance analysis, however, we made
three checks for consistency.

The first test assesses whether different authors using different data
reduction algorithms and software generated mutually consistent reduced
spectra.  The differences of equivalent widths measured by a given technique
for author A's and author B's spectra were 
$\langle W_{\rm A} - W_{\rm B}\rangle$ = $+0.4$~m\AA, with
$\sigma$~=~1.3~m\AA\ for the 1997 data, and
$\langle W_{\rm A} -  W_{\rm B}\rangle$ = $-0.3$~m\AA, with
$\sigma$~=~2.6~m\AA\ for the 1996 data.  The systematic differences are
negligible and the standard deviations acceptable, being comparable with the
expected noise. (The test sequence was not repeated with 1998 data.)

The second comparison investigates whether the net effect of using separate
reduction routes and two distinct measurement techniques is consistent with the
noise model.  The error distribution inferred from the difference between each
pair of measurements should be {\it narrower} than that based on photon noise,
since the techniques differ in the way they measure a noisy spectrum, but
sample the {\it same} data and thus are exposed to the {\it same} noise. The
measurement pairs can therefore be inspected to see whether they provide
evidence that the random noise model is optimistic.  The null hypothesis is
that the error distribution inferred from the measurement pairs is not wider
than that calculated from the model.  The error distribution of each pair of
measurements was estimated as the sample standard deviation $s_w =
{{1}\over{\sqrt{2}}}|W_{\rm Gauss} - W_{\rm sum}|$, and a standardised
statistic $Z_{\rm pair}$ was computed by dividing by the model error,
$\sigma_W$.  Only seven of the forty pairs have $Z_{\rm pair}$ values exceeding
1.0, the maximum value being 1.8, so the null hypothesis could not be rejected.
That is, we sought and failed to find evidence that the noise model is
optimistic and should not be trusted.

The third test was to compare equivalent widths we measured from spectra
obtained on more than one epoch, to check for repeatability.  We began by
computing the variance$^{-1}$-weighted mean equivalent width for each star,
$\bar{W}$, and the variance of the weighted mean, $\sigma_{\bar{W}}$ (e.g.
Bevington 1969).  These values are given in Table~2.  Next, we computed the
standardised residual, $Z$, for each observation as $Z = (W -
\bar{W})/\sigma_W$. The $Z$ distribution has a standard deviation of 1.0 (or
1.1 if restricted to stars with three observations),	
standardised residual with the largest magnitude is +2.1, which shows that all
but one of the thirty values falls within $\pm 2$ 
standard deviations of the mean. In other words, the repeatability achieved
from run to run is again consistent with the noise model.

\subsubsection{External Errors}

Although it is internal consistency which is most important for this study, it
is nevertheless valuable to know whether or not our data are consistent with
the work of others.\footnote{We elected not to risk decreasing the homogeneity
of the data set by combining it with other studies from the literature,
including our own earlier work.} Norris et al. (1994) and Ryan (1995)
highlighted differences between Li equivalent width measurements for LP~815-43,
which ranged over a factor of two from 13$\pm$2~m\AA\ and 15$\pm$3~m\AA\
(Norris et al. 1994) to 22$\pm$2.1~m\AA\ (Thorburn 1994) and
27$\pm$(3--6)~m\AA\ (Spite \& Spite 1993).  Our new measurement,
16.1$\pm$1.6~m\AA , is consistent with our earlier measurements, and
reemphasises the importance of homogeneity in obtaining small {\it random}
errors. It is the development of a large {\it homogeneous} data set in the
current work which has allowed us to probe the scatter about the Li plateau
with a much higher precision than previous work, typically conducted at the
0.06--0.10~dex level.

Most studies in which new Li data have been presented have analysed only a
dozen or so stars, so it is difficult to establish what, if any, systematic
differences exist for the various studies, as the systematics can differ from
one study to the next. Ryan et al. (1996a) addressed this issue using the
extensive data set of Thorburn (1994) as a baseline for comparison, but found
``either there were too few stars for a reliable comparison, or else the
differences that existed could not confidently be ascribed to systematic errors
amenable to transformation'' onto a unified system.  The one exception was a
small but clearly systematic offset (3~m\AA) for the Thorburn vs Spite \& Spite
(1993) samples.

In Table~3 we present previous Li equivalent width measurements of our program
stars and the three ``standard'' stars.  Perusal of the list shows no cause for
alarm that our data are systematically different from our previous work or that
of others, except perhaps for the Spite \& Spite (1993) sample as discussed
above. The most precise observations in the table are the high S/N, high
resolving power data obtained by Smith, Lambert, \& Nissen (1998) (using
different facilities to us) to measure the $^6$Li/$^7$Li isotope ratio.  We
plot our measurements against theirs in Fig.~2. For the five stars in common,
two agree within 1$\sigma$.  and the remaining three agree within
1.2--1.8$\sigma$.  This comparison leaves us confident that, even though it is
the high internal precision that is required for this study, our equivalent
width measurements are also of high absolute accuracy.

\subsection{Radial Velocity Measurements}

To provide a check on unrecognized binarity among our program stars, we have
also measured precise radial velocities.  Although there are few spectral lines
near Li~6707, the \'echelle spectra extend sufficiently blueward to include the
Mg `b' triplet and neighboring lines. The spectra were cross-correlated over
the wavelength region 5160--5200~\AA, using the 1997 observation of HD~140283
as the template. The zeropoint velocity was then set by measurements of 42
apparently unblended lines in that spectrum, which gave a formal error of $\pm
0.1$~km~s$^{-1}$ (1 s.e.).

The heliocentric radial velocity for each epoch is given in Table~4, along with
measurements from Carney et al. (1994 --- CLLA94 in the table). The ``Notes''
column gives, for the Carney et al. entries, the dispersion ($1\sigma$) of
their velocity measurements, the number of observations made, and the span (in
days) of their series of observations. Excluding the previously known
single-lined spectroscopic binary (SB1) --- BD+20$^\circ$2030 --- and one clear
new detection in this work, CD$-71^\circ$1234, the typical scatter for our
multiple measurements and for the difference between our measurements and those
of Carney et al. is 0.3~km~s$^{-1}$ (1$\sigma$). This is consistent with the
external accuracy we have obtained previously with similar observational
material (Norris, Ryan, \& Beers 1997).  There is no overwhelming evidence for
binarity in the other stars at this level of accuracy; unrecognised binaries
must have very low velocity amplitudes and/or very long periods which,
statistically at least, suggests that their companions will have minimal impact
on our analysis.  Consequently, we may infer that the impact of unrecognised
binarity is minor.

\section{Effective Temperatures}

\subsection{Observational Indices}

Deliyannis et al. (1993) attempted to circumvent the uncertainties in
color-temperature transformations by working with color alone.  However, it is
implicit in such a procedure that the color, along with its random errors,
accurately ranks the stars over the full range of the sample.  We have taken a
different approach to minimise the effects of errors, that of restricting the
diversity of stellar types at sample selection.  The dereddened
$c_1^0$~vs~$(b-y)_0$ diagram (Fig.~3) confirms that the stars are within
0.05~mag in $b-y$ of the Population II main sequence turnoff.  Nevertheless, to
reach the desired level of accuracy we need to resolve even small temperature
differences between almost identical stars, and hence fully utilise the
available temperature indices (e.g. Spite et al. 1996).  We can improve on the
studies that adopt only a single color by having up to six indices ($b-y$,
B--V, V--R, R--I, $\beta$, and HP2) on which to base effective temperatures;
these are shown in Fig.~4 as a function of $(b-y)_0$.

Several color-effective temperature scales from the literature are also shown
in Fig.~4.  Panel (a) shows the B$-$V vs $b-y$ theoretical colors of Bell \&
Oke (1986) for [Fe/H]~=~$-$2 and lg~$g$~=~4.0, which coincide roughly with the
data, and the transformation of Magain (1987, eq(15) \& (16)) at
[Fe/H]~=~$-2.8$ (the mean metallicity of our sample) which sits away from the
data, showing that Magain's B$-$V and $b-y$ scales are not mutually consistent
for these stars.  Magain's and Bell \& Oke's $b-y$ scales are almost identical
for metal-poor turnoff stars, Bell \& Oke's scale being hotter by 7~K at 6100~K
and 22~K at 6300~K.  The Bell \& Oke  (V$-$R)$_C$ and (R$-$I)$_C$ colors are
shown in Figs 4(b) and 4(c).

Panels (d) and (e) show the good correlation between $(b-y)_0$ and the Balmer
line indices, $\beta$ and especially HP2. A least-squares fit to the data
permits estimates of the $(b-y)_0$ colors from the observed Balmer indices,
which we call $(b-y)_\beta$ and $(b-y)_{\rm HP2}$.

\subsection{Calibrations}

As our sample spans a 1~dex range in metallicity, it is important to understand
the sensitivity of the effective temperature indicators to [Fe/H].  In very
metal-poor turnoff stars, we do not expect the chosen indices to be sensitive
to abundance. We sought to verify this through available effective temperature
calibrations, and especially to check the sensitivity of B--V since this index
was expected to have the greatest dependence, if any.

According to Magain's (1987) B$-$V empirical calibration, which is based on 11
stars with [Fe/H]~$<$~$-1$ and temperatures derived from the infra-red flux
method (IRFM), changing [Fe/H] from $-2.5$ to $-3.5$ would change the inferred
$T_{\rm eff}$ of turnoff stars by less than 1~K. However, the small size of
this figure may be driven by the analytical form of Magain's fitting function,
which is linear in metallicity ($Z/Z_\odot$) and hence loses sensitivity to
this variable at even moderate metal deficiency. It also should be borne in
mind that the most metal-poor member of Magain's calibration set was HD~140283,
for which [Fe/H]~=~$-2.6$ (Ryan, Norris, \& Beers 1996b), at the upper end of
the present sample's metallicity range.

One recent and extensive calibration is the table of synthetic colors computed
by Kurucz (1993) for a comprehensive range of observable indices, tracing
metallicity sensitivity down to [Fe/H]~=~$-5$.  Gratton, Carretta, \& Castelli
(1996) have shown that zeropoint differences are still found with other
calibrations, but they nevertheless adopted the Kurucz metallicity dependence
in devising their own transformation.  This metallicity dependence is shown in
Fig.~5(a)-(e) ({\it solid curves}) for pairs of colors typical of metal-poor
turnoff stars.  Magain's B$-$V and $b-y$ calibrations are also shown ({\it
dashed curves}).

A more recent empirical calibration is that by Alonso, Arribas, \&
Martinez-Roger (1996a), which uses the large calibrating set of IRFM
temperatures of Alonso, Arribas, \& Martinez-Roger (1996b).  They give fitting
functions for a wide range of stellar types, but unfortunately these become
non-physical for turnoff stars below [Fe/H]~$\sim$~$-2.5$. Figure~5(a) (dotted
curves) shows the run of $T_{\rm eff}$ as a function of [Fe/H] for a pair of
B--V colors (0.35 and 0.40) appropriate to turnoff stars.  Although the curves
exhibit a reduction in sensitivity to metal abundance as [Fe/H] falls from
$-1.0$ to $-2.6$, the fitting function goes through a minimum and climbs again
at lower metallicity.  It is unreasonable to expect that stars of yet lower
[Fe/H] exhibit {\it stronger} sensitivity to metallicity, and such behavior is
not supported by the Kurucz colors.  Clearly this behavior reflects the form of
the fitting function and the values of the coefficients, rather than the
characteristics of metal-poor turnoff stars. (Alonso et al's fitting functions
have a quadratic form so do not have the monotonically decreasing property of
Magain's.) Although Alonso et al's B--V calibration may be an improvement for
the majority of stars, it is not applicable to our very metal-poor sample.
Similar results obtain for most other indices in Alonso et al's calibrations.
Figures 5(b)-(e) show that the V--R, $b-y$, and $\beta$ calibrations also show
non-physical forms at very low metallicity, $b-y$ being the most dramatic.  We
have to disagree with Alonso et al. that their transformation equations are
valid down to [Fe/H]~=~$-3.5$.  The most we can infer is that the metallicity
sensitivity appears to saturate (reaches a minimum) by the time [Fe/H] falls to
$-2.5$, and for some indices saturates at considerably higher [Fe/H].

The metallicity range of our sample is indicated by a solid bar in Fig.~5(a).
What is important for the present study is that {\it over the metallicity range
of our sample, $-3.5 \le $~[Fe/H]~$\le -2.3$, all of the effective temperature
indices we have used should possess essentially zero sensitivity to
metallicity}. In Kurucz's calibration, which is the only one of the three
sensitive over our abundance regime, no index demonstrates a change by more
than 18~K over the interval from $-3.5$~$\le$~[Fe/H]~$\le$~$-2.5$.  We thus
feel justified in assuming that there is no significant metallicity dependence
in any of our photometric indices.

We note for completeness that we inspected the data for any metallicity
dependence in the difference between the dereddened B$-$V color and that
predicted from $b-y$ by Bell \& Oke's B$-$V vs $b-y$ calibration.  No
significant dependence was found.

\subsection{Combining Indices}

We calculated temperatures for each of the indices shown, following Magain and
Bell \& Oke in adopting linear relationships between $T_{\rm eff}$ and color
over the short temperature range involved, and adopting zero sensitivity to
metallicity due to the considerable metal-deficiency of our sample.  Minor
extrapolation was required to use the Bell \& Oke calibrations for stars hotter
than 6250~K.

Temperature scales from different indices are seldom in agreement.  We went
through the exercise of computing linear transformations between the different
temperature scales, but given the short temperature baseline covered by our
stars we doubted the reliability of the scale factors (slope coefficients), and
have instead applied zeropoint adjustments only.  We use Magain's $b-y$ scale
as the zeropoint (which essentially matches Bell \& Oke's $b-y$ scale), and
offset the Bell \& Oke-scale temperatures as follows:
$T$(V--R)$~=~T$(V--R)$_{\rm BO} - 165$~K; $T$(R--I)$~=~T$(R--I)$_{\rm BO} -
155$~K; and $T$(B--V)$~=~T$(B--V)$_{\rm BO} - 85$~K.  We also computed a
temperature (on Magain's scale) based on the estimates $(b-y)_\beta$ and
$(b-y)_{\rm HP2}$.

The adopted effective temperature for each star is the variance$^{-1}$-weighted
mean of the $b-y$, $(b-y)_{\beta}$, $(b-y)_{\rm HP2}$, and re-based B--V, V--R,
and R--I temperatures, using the variances for the individual temperature
estimates determined from the photometric errors.  The error estimates in
temperatures derived from the $\beta$ and HP2 indices include both the
uncertainty in the measurement of each spectral index itself and the
uncertainty in the $(b-y)_0$ value that is inferred from the least-squares fit
(Fig.~4(d) and (e)).  Temperatures and uncertainties are given in Table 2 (to
the nearest 10~K).  Temperatures from B--V, $b-y$, $\beta$, and HP2 are
available for almost all stars, resulting in an average over four estimates
(though the errors in $T_\beta$ result in low weight for the $\beta$ index),
with additional data from V--R and R--I being available for roughly half of the
sample.

\section{The Observed and Intrinsic Spreads in Lithium Abundance}

Figure 6(a) presents the lithium equivalent widths, $\bar{W}$, as a function of
effective temperature.  As the Li line is weak, its equivalent width varies
linearly with abundance, so lg~$\bar{W}$ is linear in logarithmic abundance.
In Fig.~6(b), the solid curve corresponds to the lithium abundance $A({\rm Li})
= 2.11$, based on the computation using Bell models at lg~$g$~=~4.0 and
[Fe/H]~=~$-2$ presented by Ryan et al. (1996a, Table~5).  Individual abundances
are shown in Fig.~6(c).  The ultra-Li-weak star G186-26 is not shown in these
figures.

The random error in each abundance measurement is taken to be the quadratic sum
of the components due to errors in W and in the estimated temperature,
$$\sigma^2_{\rm err} = ({{\partial A}\over{\partial lg W}})^2
\sigma^2_{lg W} + ({{\partial A}\over{\partial T}})^2 \sigma^2_T $$

\noindent 
Although we compute the error for each star individually, it is useful to make
a general estimate of $\sigma_{\rm err}$ for the ensemble using mean values
from Table~2: $\langle {\bar{W}} \rangle$~=~21~m\AA, $\langle \sigma_{\bar{W}}
\rangle$~=~1.3~m\AA, and $\langle \sigma_T \rangle$~=~32~K.  Since ${{\partial
A}\over{\partial T}}$~=~0.00065~dex~K$^{-1}$ for turnoff stars, we expect
$\sigma_{\rm err}~\simeq~0.033$~dex.

The standard deviation of the 22 observations is $\sigma_{\rm
obs}$~=~0.053~dex. 	
the dispersion found by Thorburn (1994), who viewed the dispersion as
significant, and by Spite et al. (1996) and Bonifacio  \& Molaro (1997), who
claimed the dispersion was within their errors. The study by Deliyannis et al.
(1993) noted a range of possible dispersions depending on the composition of
the sample.

The value $\sigma_{\rm obs} = 0.053$~dex            	
includes the contributions of quantifiable uncertainties in the data. We also
need to consider the possibility that we have observed an admixture of stellar
types not purely representative of Li plateau stars.  On the second point, a
striking feature of the observations is that the vast majority of the stars, 20
of the 22 measurements, fall within 0.1~dex of the mean.  The form of the
distribution is shown in Fig.~6(d), as both a histogram and a stripe plot, the
latter avoiding the undesirable effects of binning.  These suggest a roughly
Gaussian distribution about the mean with a small dispersion, plus two stars
lower in abundance by $\simeq$0.14~dex.  Indeed, the standard deviation for the
20 stars within $\pm$0.1~dex of the mean is a mere $\sigma_{\rm
obs}$~=~0.036~dex.  The two stars with lower abundances therefore represent
4.1$\sigma$ (CD$-24^\circ$17504) and 3.4$\sigma$ (BD+9$^\circ$2190) deviations,
address the reasons they differ below, but for now we emphasise that for 20 of
the 23 stars in Table~2, i.e. for 87\% of the very metal-poor, halo-turnoff
sample, the observed spread in Li abundance is only $\sigma_{\rm
obs}~=~0.036$~dex.  Since we estimated the random error for the ensemble to be
$\sigma_{\rm err}~=~0.033$~dex, it is clear that the vast majority of the
sample is consistent with essentially zero scatter about the mean. In other
words, {\it the Spite Li plateau is ultra-thin at the metal-poor turnoff.}

Three stars are highlighted in Fig.~6 and Fig.~7, where the latter shows the
scatter about the Li plateau in standardised units $Z_i = (A({\rm Li})_i -
2.11)/\sigma_{A({\rm Li})_i})$.  Two of the stars were introduced above.
BD+9$^\circ$2190 falls well below the mean, but by only 2.8 times its formal
error, so its position could be consistent with its errors.  In contrast,
CD$-24^\circ17504$ and CD$-71^\circ1234$ lie away from the mean by
(respectively) 4.7 and 3.5 times their formal errors, suggesting that they have
genuinely different Li abundances from the rest.  In \S4.2.3, we searched for
but failed to find evidence that the formal error estimates were unreasonable;
it would be ad hoc, without more evidence, to suggest that these two stars are
exceptions, especially since each has several observations.  Even exclusion of
CD$-24^\circ$17504's most extreme datum --- 1997: $W = 15.1$~m\AA\ --- would
leave the star below the mean by 3.0 times its (revised) formal error.  We
conclude that both stars lie significantly away from the mean, and discuss the
cause of this below (\S 7.3.1).

\section{Discussion}

\subsection{Comparison with Previous Measurements of Spread}

The essentially zero spread found for the Li plateau at the very metal-poor
turnoff may be contrasted with larger values found in several previous studies.

Spite et al. (1996) and Bonifacio \& Molaro (1997) both considered the spread
they measured to be consistent with zero to within their formal errors.  Our
new result offers no contradiction, but due to the much better precision
achieved in our study, $\sigma_{\rm err} \simeq 0.033$~dex compared with
0.06--0.08~dex (Spite et al.) and 0.07~dex (Bonifacio \& Molaro), our result
can be stated much more strongly.  Our better precision derives from the use of
a very homogeneous data set, the checks undertaken to ensure that error
estimates were appropriate (e.g. through double-blind processing and double
measuring of every spectrum), the utilisation of multiple indices to minimise
random errors in effective temperature, and the application of restrictive
selection criteria which minimised physical differences between the stars.
Otherwise, star to star differences might have induced greater temperature
and/or metallicity dependent errors associated with color-effective temperature
transformations and model atmospheres. We cannot claim that our abundance
calculations have completely overcome the systematic errors --- we quantify
them in \S7.7 --- but we have avoided them insofar as they affect measurements
of the thickness of the Li plateau.

Given the essentially zero intrinsic scatter found for our sample, how are we
to interpret the earlier measurements of significant scatter by Deliyannis et
al. (1993) and Thorburn (1994)?  We reexamine these studies in reverse
chronological order.

Thorburn (1994) acquired an almost homogeneous data set, making extensive
observations (utilising four different telescope/spectrograph combinations) and
quoting formal errors in the range 0.08--0.09~dex. Although the bulk scatter in
abundances was 25\%, this reduced to 15\%\ once $T_{\rm eff}$ and [Fe/H] trends
were removed. It is this latter figure which is relevant to the thickness of
the Li plateau.  Thorburn noted that the total formal error would have to be
increased by $\sim$20\%\ to explain the observed scatter, 
$\simeq$~0.1~dex, and suggested that the scatter may be a consequence of
dispersion in the halo age-metallicity relationship and Galactic chemical
evolution.  However, the much smaller scatter we have found, $\simeq$~0.03~dex,
obviates the need for such an explanation.  We cannot be certain of the reason
for the excess scatter in Thorburn's study, but we argue as follows that it may
be artificial.  Four different instrumental setups were used, but neither sky
nor scattered light subtractions were made, which Thorburn estimated could
introduce errors of not more than 1--2\%\ and possibly 3--5\%\ respectively. It
is conceivable that differences between the scattered light and sky backgrounds
from telescope to telescope and from night to night have contributed to the
scatter in the data.  >From Thorburn's Table~2, the formal $\sigma(W)$ is
typically only 10\%\ of $W$. The errors from neglect of sky and scattered light
will contribute 0\%\ of $W$ in the optimistic case and 7\%\ of $W$ in the
pessimistic case, so the actual errors should be higher than the stated values
by between 1.0 and 1.7 times.  It is conceivable that the stated $\sigma(W)$
values do underestimate the actual errors in $W$ sufficiently to explain the
1.2 times higher than expected scatter.

Deliyannis et al. (1993) studied an inhomogeneous compilation of data from the
literature, and quantified the uncertainties of each measurement using a noise
model of the type discussed above (e.g. Cayrel 1988). They considered a number
of subsamples, and found a dispersion of $\pm \ge$20\%\ (2$\sigma$), i.e.
$\sigma \ge$~0.04~dex, depending on which subsample was examined. This
dispersion is not much different from our observed scatter, but as noted
already, our formal errors are also at this level, so we infer $<$0.02~dex
intrinsic scatter. They computed the scatter in each sample at uniform $b-y$
color, which can be viewed as removing trends in $T_{\rm eff}$ but not trends
in [Fe/H]. Claims of a dependence of $A({\rm Li})$ on [Fe/H] had not been
published at the time of the Deliyannis et al. work. With the benefit of
hindsight, we might expect that the Deliyannis et al. scatter measurements
could be inflated by the presence of such a trend, if it exists. (We will
return to that point below.) Subsequently, Thorburn (1994) estimated the [Fe/H]
dependence of the A({\rm Li}) trend as 0.13~dex~per~dex; Ryan et al. (1996a)
derived a similar value, 0.11. The Deliyannis et al. sample ranged from $-3.5
\le$~[Fe/H]~$\le -1.4$, which would span 0.25~dex in $A({\rm Li})$ if the slope
noted above were correct. A normally distributed sample has a standard
deviation $\sim$ 1/6 to 1/4 of its range, so a sample spanning 0.25~dex might
well be expected to yield a standard deviation of 0.04--0.06~dex. Thus the
scatter derived by Deliyannis et al. is consistent with published values of the
embedded metallicity dependence of $A({\rm Li})$ and with the abundance range
of their sample.

This explanation of Deliyannis et al's findings would fail, however, if the
metallicity trend did not exist, as Bonifacio \& Molaro (1997) concluded.  We
revisit this below (\S7.3).  Bonifacio \& Molaro also pointed out that both
Thorburn's and Deliyannis et al's work used straight line fits to the data in
determining the scatter, whereas exponential fitting functions may have been
more appropriate.  Although theoretically a non-linear form may have been
better suited, it is not clear quantitatively whether the difference can be
explained in this fashion.

\subsection{Reexamination of G64-12, G64-37, and CD$-$33$^\circ$1173}

Ryan et al. (1996a) drew particular attention to G64-12, G64-37, and
CD$-33^\circ1173$ as three stars having essentially identical atmospheric
parameters but irreconcilable lithium abundance determinations. All three stars
are included in the present study, and as the conclusions already stated
indicate, we no longer identify a significant spread amongst this set of stars.
The effective temperatures (Table~2) are still within a total range of 30~K,
and the metallicities (Table~1) are within 0.10~dex. However, the homogeneous
Li equivalent widths we have measured in this work differ considerably from
those in the heterogeneous compilation of Ryan et al.  The new vs old values
(in m\AA) are respectively: G64-12, 21.1$\pm$1.1 vs 27$\pm$1.8; G64-37,
18.2$\pm$1.5 vs 15$\pm$1.0; and CD$-33^\circ1173$, 17.2$\pm$1.2 vs 12$\pm$1.2.
The formal errors in the current work differ little from the 1996 compilation,
but in view of the homogeneity which we have achieved in the new data set, we
prefer the newer measurements.  We have no detailed explanation for the
discrepancy other than to repeat the cautions given in Ryan et al. (1996a) and
elsewhere that it is easy to overlook or misjudge error contributions when
making error estimates and combining heterogeneous data sets.

\subsection{Examination of A({\rm Li}) vs [Fe/H]}

\subsubsection{Morphology of the Sample}

Trends of $A({\rm Li})$ with both $T_{\rm eff}$ and [Fe/H] were cited by Norris
et al.  (1994), Thorburn (1994), and Ryan et al. (1996a).  However, Bonifacio
\& Molaro (1997) concluded that these were eliminated by using the IRFM
temperatures of Alonso, Arribas, \& Martinez-Roger (1996b).

We chose our sample to be very metal-poor, both to minimise the differences
between stars in the study and to obtain measurements of objects which show the
least signs of chemical enrichment. However, the stars do span a small range of
metallicity and, given the proven accuracy of the data, are useful for
examining again the metallicity dependence of $A({\rm Li})$.  We plot Li
abundance vs [Fe/H] in Fig.~8(a).  Recall from \S3 that the [Fe/H] values are
mostly based on high and/or medium-resolution spectroscopic observations, for
which $\sigma_{\rm [Fe/H]} \simeq 0.15$~dex.  It is clear at first glance that
a similar trend with [Fe/H] is identified in the present study as was measured
by Thorburn (1994) --- 0.13~dex per dex --- and by Ryan et al. (1996a) ---
$0.111 \pm 0.018(1\sigma)$.  An ordinary least squares (OLS) fit, excluding
only G186-26, gives d$A$(Li)/d[Fe/H]~=~0.121($\pm$0.028) (errors are standard
errors), with a scatter about the trend $\sigma_{\rm obs}$~=~0.037~dex (dotted
line, Fig.~8(a)).  Moreover, although we discussed above  whether
CD$-24^\circ17504$ and CD$-71^\circ1234$ should be included or excluded, it is
clear that they lie on the {\it same} trend as the rest of the data in
Fig.~8(a).  This justifies our confidence in the quality of the observational
data.

Because of possible concern whether the trend is real or illusory, we undertook
a series of regression analyses, excluding a priori G186-26.  These included
ordinary least squares (OLS), reweighted least squares (RWLS --- Rousseeuw \&
Leroy 1987), which is a robust technique that detects outliers, the
BCES\footnote{Bivariate Correlated Errors and intrinsic Scatter (Akritas \&
Bershady 1996)} approach adopted by Bonifacio \& Molaro (1997), which has
regard for the various error contributions in each datum, and a robust
technique based on a bisquare regression procedure described by Li (1985).  The
first step was to undertake OLS and RWLS bivariate analyses of
$A$(Li)~=~$a_0~+~a_1\times$[Fe/H]$~+~a_2\times T_{\rm eff}$.  Detailed results
are presented in Table~5. The coefficient of determination, $R^2$, listed in
the table indicates the proportion of variance in the dependent variable which
is explained by the independent variable(s) in the regression model.  For both
techniques, and also for culled subsets of the data, we found the coefficient
of $T_{\rm eff}$ to be indistinguishable from zero, to a precision of
$\simeq$~0.010 (1$\sigma$) dex per 100~K.  This is not {\it entirely}
surprising given the short temperature interval for the data, but is
nevertheless inconsistent ($>$3$\sigma$) with the result of a previous analysis
of heterogeneous data (Ryan et al. 1996).  The bivariate RWLS analysis rejected
BD+09$^\circ$2190 as an outlier, but this did not alter the redundant status of
the $T_{\rm eff}$ coefficient.  Whatever the explanation for the difference
between the Ryan et al. (1996) sample and the current one, clearly a
temperature term is unnecessary in the present analysis, and all further tests
were conducted using univariate fits of the form $A$(Li)~=~$a_0 + a_1$[Fe/H].

The OLS univariate fit to the data is shown as a dotted line in Fig~8(a).  The
RWLS fit again identified BD+9$^\circ$2190 as an outlier, and the fit to the
remaining stars is shown with the solid line. This represents our ``best fit''
--- $$A{\rm (Li)}~=~2.447(\pm0.066)~+~0.118(\pm0.023)\times{\rm [Fe/H]}.$$ (The
RWLS regression is identical to that which would be obtained from the OLS fit
if BD+9$^\circ2190$ was excluded a priori.) Clearly, the result is barely
sensitive to the inclusion or exclusion of this star. The coefficient of [Fe/H]
is found to be non-zero at a high significance, viz.
0.118$\pm$0.023(1$\sigma$).  The same conclusion was reached from the
assortment of other regression tests preformed (see Table~5 for details).

Earlier in the discussion, we identified CD$-24^\circ$17504 and
CD$-71^\circ$1234 as deviating from the mean by more than their formal errors.
This can now be understood in terms of their rankings at the low and high end
of the metallicity scale. Although they were not identified as outliers by the
RWLS fit, we considered further the possibility that they might carry excessive
weight in influencing the trend, and conducted tests on a culled sample. A RWLS
regression from which BD+9$^\circ$2190 was culled a priori subsequently
identified CD$-24^\circ$17504 and  CD$-71^\circ$1234 as outliers, and gave a
shallower, but still significantly non-zero, slope for the trend --- $$A{\rm
(Li)}~=~2.318(\pm0.063)~+~0.073(\pm0.022)\times{\rm [Fe/H]}.$$ The fit for the
culled sample is shown in Fig.~8(e).  Alternative regression fits for this
sample are given in Table~5; all give significantly non-zero values for the
slope.\footnote{We note for completeness that the star HD 74000, which fell
outside the metallicity and temperature range of our sample selection criteria,
was nevertheless observed as a standard star, in order that we could compare
our equivalent width measurements with those of other workers. We noticed,
however, that it has a lower $A$(Li) abundance than most other stars in our
sample. If it had been included in the target group, it too would have been
rejected by the outlier--detection routines in our regression analyses. If
$^7$Li is genuinely depleted in this star, this may account for the
non-detection of $^6$Li despite it being only 100~K cooler than HD~84937 and
BD+26$^\circ$3578 in which $^6$Li is seen (Smith et al. 1998).}

Figure~8(b) and 8(f) give the residuals of $A$(Li) about the regression
functions, with open circles indicating data excluded from the fit.  Histograms
and stripe plots (Fig.~8(c) and 8(g)) show the residual distributions.  The
``best fit'' yields a dispersion $$\sigma_{\rm obs}~=~0.031~{\rm dex},$$ while
the ``culled fit'' has $\sigma_{\rm obs}$~=~0.024~dex.  Robust biweight
estimators of scale (see Beers, Flynn \& Gebhardt (1990) and references
therein) yield values $S_{BI}$~=~0.031 and 0.025 respectively.  (The biweight
estimate of scale converges to the standard deviation estimator when sampling
from a normal distribution, but is less sensitive to the presence of outliers).
The normality of the $A$(Li) residuals is established not only by the excellent
agreement between the $\sigma$ and $S_{BI}$ values, but also via the lack of
departure from linearity  in the ``normal probability plots\footnote{A normal
probability plot ranks the data from lowest to highest, and plots the ordered
value against its theoretical Z-statistic. The Z-statistic gives the number of
standard deviations by which the datum would depart from the mean in a normal
distribution of $N$ points.  For a normal distribution, ranked datum $i$ will
possess probability value $i/(N+1)$ measured from the lowest tail, so the
cumulative probability distribution is inverted to find the corresponding
Z-statistic. If a data set is normally distributed, then it will lie about a
straight line in the plot, whereas an asymmetric distribution will deviate from
the line along a curved path.  See, e.g., Levine, Berenson \& Stephan
(1998).}'' in Fig.~8(d) and 8(h).

The regression analysis may be summarised thus: We have found a positive
dependence of $A$(Li) upon [Fe/H] (but not $T_{\rm eff}$) which resembles the
values found previously by Thorburn (1994) and Ryan et al. (1996). Our best fit
gives d$A$(Li)/d[Fe/H]~=~0.118($\pm$0.023).  Shallower values of the slope can
be obtained by a priori rejection of some of the data --- which may be an
invalid action --- leading to d$A$(Li)/d[Fe/H]~=~0.073($\pm$0.022), but even
then the slope is significant at $\ge$3$\sigma$. The scatter measured for the
best fit is $\sigma_{\rm obs}$~=~0.031~dex.  Obviously, rejection of stars to
obtain a shallower slope yields even smaller values of the scatter, but in any
case the observed scatter is consistent with the expected errors $\sigma_{\rm
err}$~$\simeq$~0.033.

\subsubsection{Is the Trend Natural or Artificial?}

A major similarity between the Ryan et al. (1996a) study and the present one is
the use of the same computations relating equivalent width to abundance.  If a
metallicity-dependent error existed in that work, it would persist here. Can
such an error be identified?

The Ryan et al. (1996a) work used model atmospheres from Bell (1983), computed
at [Fe/H]~=~$-$2. If the model structure differed sufficiently for real stars
between [Fe/H]~=~$-$3.0 and $-$2.0, a metallicity dependent error might be
expected. However, the Kurucz (1993) models (which extend to lower abundance
than Bell's but have high convective overshoot) show that changing from the
higher to the lower metallicity would change the inferred Li abundance by only
0.012~dex (see Ryan et al. 1996a, Fig.~2).  This is an order of magnitude less
than the trend identified and a factor of three smaller than $\sigma_{\rm
obs}$, and in the sense of steepening rather than flattening the trend. On this
estimate, the atmospheric models are not sufficiently sensitive to metallicity
to produce the trend we observe.  Note also that Thorburn used Kurucz (1993)
models rather than the Bell models as adopted here, yet derived an almost
identical trend. This emphasises again that selection of a different model grid
may alter the derived absolute abundance, but will have little effect on the
differential characteristics of the results.

An alternative source of error might be a metallicity dependence in the
effective temperature scale. Our effective temperatures were based
substantially on B--V, $b-y$, and HP2.  In \S5 we argued that all of the
indices used are insensitive to metallicity for very low-metallicity turnoff
stars.  An error of 100~K in effective temperature would produce an abundance
error of 0.065~dex for stars of the temperature and metallicity of our sample,
so an error of 200~K would have to be induced {\it over the short metallicity
interval from [Fe/H]~=~$-2.3$  to $-3.5$} to produce the trend observed, yet we
identified at worst an 18~K change in the Kurucz (1993) color transformations.
In view of the lack of sensitivity of our temperature indicators for the types
of stars investigated, we do not believe that the trend can be explained away
in this fashion.

We have ruled out metallicity-dependent errors in the stellar atmospheres and
effective temperature scales as causes of the trend. We do not expect such
errors in the equivalent width measurements either, since the spectra are
devoid of lines around Li~6707~\AA\ and the continuum fit should be reliable
irrespective of metallicity for our stars. NLTE effects were assessed and
rejected as the cause by Ryan et al. (1996a).  We are left with little
alternative but to restate our identification of the trend over the interval
$-3.6~<$~[Fe/H]~$<~-2.3$, and to consider it to be natural until proved
otherwise.

\subsubsection{The Bonifacio \& Molaro Analysis in Retrospect}

The metallicity dependence derived here is very similar to that found by Ryan
et al. (1996a) and Thorburn (1994), but the new sample is far more homogeneous
and of much higher quality.  How then should we view Bonifacio \& Molaro's
(1997) conclusion that there is no metallicity dependence?  Their work used
IRFM temperatures, which one might arguably prefer over other scales,
especially as far as systematic errors are concerned, but there are two crucial
disadvantages of their study compared with ours. Firstly, the formal errors in
the IRFM temperatures listed by Bonifacio \& Molaro are typically 80~K, whereas
by averaging many different indices we have reduced the random error to
typically 30~K.  The larger errors of the IRFM temperatures induce greater
random scatter about the Li plateau for that dataset.  Secondly, their
equivalent widths and [Fe/H] values were based on a literature survey of
inhomogeneous and less reliable data than in our new work.  The combined effect
of these factors is that, although their bivariate fit of $A$(Li) on [Fe/H] and
$T_{\rm eff}$ gave a slope consistent with zero, the uncertainty in its
determination was sufficiently large that our new value lies at only their
2.5$\sigma$ tolerance\footnote{Their bivariate fit for an LTE analysis without
depletion corrections (i.e.  matching our assumptions) gave a metallicity slope
0.034$\pm$0.034~dex~per~dex.}.  Bonifacio \& Molaro's univariate fits, however,
are irreconcilable with our result, having [Fe/H] coefficients ranging from
$-$0.02 to $-$0.05~dex per dex and uncertainties (1$\sigma$) of 0.03 to
0.06~dex per dex, depending on the statistical test. In what follows, we
identify additional reasons for the differences between their result and ours
obtained with the current sample.

The difference between our estimated slope of $A$(Li) on [Fe/H] and that of
Bonifacio \& Molaro can be explained upon closer scrutiny of the literature
data used in their study.  Figure~9 shows the subset of nine stars common to
both works.  We use the new homogeneous $W$(Li) values from this study, but
show abundances calculated on both the IRFM and our temperature scales (central
and upper panels respectively). We plot these against both the literature
[Fe/H] values referenced in Table~2 (left hand panels) and the values used by
Bonifacio \& Molaro (right hand panels).  The effect of using the IRFM scale is
to generate huge scatter (central panels) due to the low precision of those
individual values. As emphasised previously, high internal precision is
required to assess the spread about the Li plateau, and this precision is
delivered by the variance$^{-1}$-weighted average over three to six different
temperature indicators, not by the use of a single ``noisy'' index even if the
latter may have better systematics.

Restricting our attention, then, to the low scatter (uppermost) panels using
the temperatures computed in this work, it is clear that the trend with
metallicity depends on the adopted metallicity estimates. Without more
information, it would not be possible to know whether the literature [Fe/H]
compilation in Table~2 or that used by Bonifacio \& Molaro is better.
Fortunately, we do have more information, in the second set of [Fe/H] values
derived from applying the calibration of Beers et al. (1999) to our
1\AA-resolution spectra (see Table~2). In the bottom panels of Fig.~9, we
compare the metallicities derived from those 1\AA-resolution spectra with the
adopted literature values (Fig.~9(e)) and those used by Bonifacio \& Molaro
(Fig.~9(f)), and find excellent agreement with our adopted literature values,
but considerable disagreement with some of the values adopted by Bonifacio \&
Molaro, to the extent that the plot in Fig.~9(b) becomes levelled off by the
scatter in [Fe/H].  For completeness, we note that an OLS regression of our
presently derived $A$(Li) estimates with the Molaro \& Bonifacio values of
[Fe/H] for the nine stars in common (Fig.~9(b)) results in a slope with respect
to abundance of 0.008 (+/- 0.041), i.e., completely consistent with zero.
However, on the basis of these comparisons, we favor the literature [Fe/H]
values adopted in Table~2 to those adopted by Bonifacio \& Molaro. Preferring
also the temperatures derived from multiple indices rather than the individual
temperatures based on the IRFM, we believe that Fig.~9(a) is the most reliable
presentation of the data.

Restating our result above, allowing for the [Fe/H] dependence in our sample
with only G186-26 and BD+9$^\circ$2190 excluded, {\it we find a tiny
dispersion, $\sigma_{\rm obs}$~=~0.031~dex, for 91\%\ of the sample}.  It
remains now to discuss the significance of the trend with metallicity.

\subsection{$^6$Li as a Tracer of Non-Primordial $^7$Li}

The interpretation of halo Li abundances would be greatly simplified if the
Spite Li plateau had no dependence upon metallicity. However, we have again
measured a positive dependence. Furthermore, and even if one denies the reality
of this trend, the fact that at least two stars in our sample are contaminated
with $^6$Li indicates a distinctly non-primordial origin for {\it some} of the
Li in these stars.  Smith, Lambert, \& Nissen (1993, 1998) and Hobbs \&
Thorburn (1994, 1997) have measured the presence of $^6$Li in HD~84937 and
BD$+26^\circ3578$ at the level of $^6$Li/Li = 0.06$\pm$0.03 and 0.05$\pm$0.03
(Smith et al. 1998) respectively. Both of these stars are in our sample.  Our
data are not of high enough resolving power or S/N to measure $^6$Li
separately, but since all of the stars in our narrowly-defined sample should
have a similar evolutionary history and stellar structure, more likely than not
they will all be contaminated by $^6$Li.

In the following discussion, we assume that all of the $^6$Li is pre-stellar.
Alternative possibilities were examined by Lambert (1995), who performed an
initial appraisal of synthesis by Galactic cosmic rays stopped in the stellar
convection zone, and by Deliyannis \& Malaney (1995), who considered synthesis
by stellar flares. The former appraisal revealed potentially important
production of $^6$Li, but with large uncertainties, and on balance Lambert
viewed the mechanism as probably too inefficient. The second assessment
indicated possibly significant levels of $^6$Li production and retention in
turnoff stars, but again the calculation was subject to large uncertainties
associated with the (unknown) flare-history of the star.

As $^6$Li production at the levels measured exceeds that expected from standard
Big Bang nucleosynthesis, we infer that it originates in sources associated
with Galactic chemical evolution (GCE), and we should expect GCE production to
vary with [Fe/H]. Furthermore, since GCE $^7$Li production must accompany GCE
$^6$Li production, we have to disentangle three components to the abundances we
measure via a single spectral feature: primordial $^7$Li, GCE $^7$Li, and GCE
$^6$Li.

Ramaty, Kozlovsky, \& Lingenfelter (1996)  and Ramaty et al. (1997) give the
production ratio of $^7$Li/$^6$Li as $\sim$ 1.3--1.7 for Galactic cosmic rays
having energies and compositions consistent with Be and B synthesis. We adopt
the value 1.5 in the calculations that follow.  Assuming $^6$Li/Li = 0.05,
(i.e. $^6$Li/$^{6+7}$Li) and making the most conservative assumption that none
of the pre-stellar $^6$Li has been destroyed in these turnoff stars, we would
argue that 8\%\ of the $^7$Li, and 13\%\ of the total Li absorption in these
stars is non-primordial. We would therefore infer that the primordial value of
$^7$Li should be 0.06~dex lower than the observed $A$(Li).  If some of the
$^6$Li has been destroyed during these stars' lifetimes, as seems likely, then
the GCE $^7$Li fraction would be higher and the primordial value lower.
Destruction of $^6$Li at the turnoff is predicted to be a strong function of
mass (effective temperature) and age, and standard Yale models (e.g. Deliyannis
et al. 1990; Pinsonneault et al. 1992) show that depletion by 0.1 to 0.5~dex is
not unreasonable, and that substantially more depletion may have taken place in
practice.  If we assume that 50\%\ of the pre-stellar $^6$Li has been
destroyed, then the GCE $^7$Li component would be 17\% of the total; 21\%\ of
the current line absorption would be due to GCE, and the primordial value would
be 0.10~dex lower.

\subsection{Non-Primordial Li and the [Fe/H] Dependence}

Since the presence of $^6$Li indicates that at least some non-primordial Li is
present, it is logical to ask whether the inferred GCE $^6$Li and $^7$Li
components can explain the observed dependence of $A({\rm Li})$ on [Fe/H]. The
calculations above show that the primordial $^7$Li abundance probably is {\it
at least} 0.06~dex lower than the $A$(Li) abundance measured in HD~84937 and
BD$+26^\circ3568$, and that a value 0.10~dex lower might be a realistic
estimate. According to the measured trend, the [Fe/H] value at which $A$(Li) is
observed to be 0.10~dex lower than in the relatively metal-rich stars HD~84937
and BD$+26^\circ3568$, is [Fe/H]~=~$-3.2$.  It is important to note that we
have argued elsewhere (e.g. Ryan et al. 1991, 1996b; Ryan 1996), in work {\it
not} involving Li, that the Galaxy's first supernova enrichment events give
rise to stars around [Fe/H]~$\sim$~$-4.0$ to $-3.5$.  The metallicity
dependence we have measured for Li is therefore roughly consistent with the GCE
contribution to Li inferred up to the time when HD~84937 and BD$+26^\circ3568$
formed. It is not unreasonable to suppose, then, that the most metal-poor star
in our sample, CD$-24^\circ17504$ with [Fe/H]~=~$-3.55$, has minimal GCE
contribution to its Li line, whereas at higher metallicities we see the GCE
contribution increasing.

If $^6$Li preservation is confined to turnoff stars, as the Yale models suggest
(Deliyannis et al. 1990; Pinsonneault et al. 1992), then in cooler dwarfs we
would have to adjust only for GCE $^7$Li to obtain the uncontaminated
primordial value.  However, since GCE $^6$Li contributes less than GCE $^7$Li
to the contamination, the offset would be reduced only from 0.10~dex to
0.08~dex. This is not a major difference, but does emphasise that the
metallicity dependence may be slightly weaker in dwarfs away from the turnoff.

We have argued that the observed metallicity dependence in this very metal-poor
sample is consistent with the GCE contribution inferred from the $^6$Li
measurements in two stars at [Fe/H]~$\sim$~$-2.4$. However, if we are to claim
to understand the slope as due to GCE and hence be able to infer that the most
metal-poor stars yield the correct primordial Li abundance, then we also need
to assess whether the explanation correctly predicts the metallicity dependence
in more metal-rich stars.  Smith et al. (1998), amongst others, have noted that
if the Li/Be ratio is maintained in GCE production throughout formation of the
halo, then the Galaxy ought to have become very rich in Li by [Fe/H]~=~$-1$,
but apparently it did not. The $\alpha + \alpha$ fusion mechanism produces
roughly uniform Li throughout the phase of halo formation in contrast to the
strong metallicity dependence of Be and B (Steigman \& Walker 1992), and Olive
\& Schramm (1992, eq. (6)) predict, by comparing Li to Be, a very shallow
relationship, approximately GCE~$A(^7{\rm Li}) \simeq 1.59 + 2Z/Z_\odot$ for
$Z/Z_\odot~<~0.1$. Clearly, however, the detailed evolution of $^6$Li, and
therefore of GCE $^7$Li, also depends on the chemical evolution model adopted
(e.g. Prantzos, Casse, \& Vangioni-Flam 1993; also contrast Figs 1 and 2 of
Yoshii, Kajino, \& Ryan 1997). The lesson from these models is that the
inferred total Li abundance need {\it not} climb significantly more steeply
over the range $-2.5 < $~[Fe/H]~$<-1.5$ than it does over the interval $-3.5 <
$~[Fe/H]~$<-2.5$ which we have measured. Furthermore, since the
higher-metallicity samples often include cooler stars, the average observed
slope may flatten slightly at higher metallicity due to the erasure of the
$^6$Li contribution to $A$(Li).

In summary, we regard the slope in $A({\rm Li})$ vs [Fe/H] to be concordant
with the amount of GCE inferred from the observed $^6$Li abundances.
Furthermore, GCE models which have higher Li/Be yields at lower metallicity
(e.g. Steigman \& Walker 1992) suggest that the amount of GCE Li expected at
higher metallicities need not invalidate this explanation.  Irrespective of
whether the metallicity trend is believed (since there may be sceptics in the
readership), from the observed $^6$Li fractions we infer that the primordial
abundance is $\simeq 0.10$~dex below that with which stars having
[Fe/H]~$\sim$~$-2.4$ were born.

In addition to the galactic cosmic ray mechanism discussed above, stellar
nucleosynthesis of $^7$Li may contribute to the measured trend. D'Antona \&
Matteucci (1991) computed an increase of $A({\rm Li})$ by 0.17~dex over the
interval [Fe/H]~=~$-2.5$ to $-1.5$, for production in 2--8~$M_\odot$ AGB stars.
Although that slope is subject to uncertainties in the adopted parameters ---
they also computed models which showed steeper trends --- and could be less, it
emphasises that even at this early stage of GCE, we must recall the {\it
likelihood}, not merely the {\it possibility}, that the Li we observe in halo
stars is affected by GCE. The trend we have measured, 0.12~dex~per~dex, is
concordant with the observed $^6$Li contamination and expected stellar
production.

Before leaving this discussion, we note that the narrowness of the Li spread is
maintained over the range [Fe/H]~$<$~$-2.3$, even though GCE is leaving its
mark on material, increasing $A$(Li) and producing measurable $^6$Li. This
result sets an additional constraint on GCE models of lithium processing. One
interpretation is that any age spread in the formation of halo stars over this
low-metallicity interval must not be so great as to lead to expectations of a
measurable range of $A$(Li) at a given [Fe/H]. However, an inference on age
ranges may be relaxed in the Searle \& Zinn (1978) framework where early halo
star formation began in separate, independently evolving fragments. In these
first star formation events in the voluminous proto-halo, it is possible that
regions were sufficiently separated that cosmic rays accelerated in one part
did not reach and induce reactions in the others, so Li would evolve in concert
with the local metallicity rather than the galactic age. Unfortunately, models
of cosmic ray propagation in the voluminous proto-halo are less well
constrained than in the Galactic disk, for which we can infer present day
lifetimes, path lengths and spectra. The measurement of $^6$Li in more halo
stars will help constrain the Galactic cosmic ray production ratios
$^{6,7}$Li/Be and $^{6,7}$Li/B in the earliest phase of GCE.

\subsection{Constraints on Rotationally Induced Mixing Models}

The rotationally-induced turbulent mixing models of Pinsonneault et al. (1992)
differ from the Yale ``standard'' and ``diffusive'' models in predicting
substantial ($\sim 1$~dex) depletion of Li in halo turnoff stars. Amongst the
signatures of this depletion mechanism are a mildly-arched ``plateau'' and a
spread in final abundances reflecting the range of initial angular momenta of
the stars.

Using updated models having an improved treatment of the evolution of angular
momentum, and considering the spread in $A$(Li) seen in Thorburn's (1994) data,
Pinsonneault et al. (1998) concluded that the mean Spite Li plateau abundance
was depleted by 0.2--0.4~dex from the primordial value.  Utilising our more
accurate data for turnoff stars, we now revisit that result.

Pinsonneault et al. (1998) computed the depletion -log~$D_7$~=~$A$(Li)$_{\rm
final}-A$(Li)$_{\rm initial}$ for stars with [Fe/H]~=~$-2.3$ and $T_{\rm
eff}$~=~6000~K, coincidentally very similar to the parameters typical of our
sample. They present their results for three different solar angular momentum
histories (which affect the calibration of their models), convolved with
observational errors of 0.00 and 0.09~dex corresponding to perfect observations
and the formal error of Thorburn (1994) respectively.  As our formal errors are
only 0.033~dex, we broadened Pinsonneault et al's ``perfect'', minimal
depletion ``s0'' model by an appropriate value, and compare it to our data in
Fig.~10(a).  (Recall that G186-26 is heavily depleted and lies offscale.) The
depletion curves and data are brought into coincidence by assuming an initial
abundance $A$(Li)$_i$~=~2.22~dex.  It appears at first sight that the data are
clustered more tightly than the theoretical boundaries enclosing $\pm$47.5\%\
of the population (dashed curves).  However, it is more reliable to view the
distributions functions directly, so we have renormalized the theoretical
distribution with zero observational error (Pinsonneault et al., Fig.~9(a)) to
the number of stars in our sample, and scaled the depletion from their $T_{\rm
eff}$~=~6000~K to the mean of our sample, $T_{\rm eff}$~=6200~K.  Our raw
sample (excluding only G186-26), shown in Fig.~10(b), not surprisingly has a
broader core than the theoretical distribution because of the imbedded [Fe/H]
trend.  Fig.~10(c) overcomes the trend by shifting all stars to a common
metallicity --- [Fe/H]~=~$-2.8$, the median value of our sample --- using
$\Delta A$(Li)/$\Delta$[Fe/H]~=~0.12.  The cores of the observed and
theoretical distributions match well, but the model has a Li-depleted tail
extending to much lower abundances than the data.  Specifically, Fig.~10(c)
showing Pinsonneault et al's (1998) model ``s0'' with least depletion, predicts
that 17\%\ of the sample, or 3.9 stars for our sample of 23, will have
$A$(Li)~$<$~2.0 (at [Fe/H]~=~$-2.8$).  In fact we observe only one star below
this limit, G186-26, and even that is excessively depleted compared to the
model, as if some factor other than the Pinsonneault et al. mechanism is
responsible.  We conclude that even the minimally depleting ``s0'' model of
Pinsonneault et al. overpredicts the degree of Li depletion in the turnoff
stars. Whereas the Thorburn sample allowed Pinsonneault et al. to infer
depletion by 0.2--0.4~dex by this mechanism, the higher quality data now
available give rise to two new conclusions: (1) even the ``s0'' rotational
model with a median depletion as small as 0.1~dex at $T_{\rm eff}$~=~6200~K
predicts a broader spread than permitted by the turnoff observations; and (2)
the very low Li abundance in G186-26 is not consistent with the
rotational-depletion distribution function. The latter result signals that this
star, and consequently the other ultra-Li-depleted halo dwarfs, do not
represent the tail of a rotational depletion distribution.  It is no longer
possible to infer a minimal rotational depletion of 0.2~dex as Pinsonneault et
al. were led to do with less accurate data.

The extremely tight clustering of the halo turnoff stars therefore presents a
serious challenge to inferences from {\it this} class of models that the
turnoff stars have depleted by even as little as 0.1~dex from a higher initial
value.

\subsection{The Primordial $^7$Li Abundance}

Several estimates of the primordial Lithium abundance, $A$(Li)$_p$, can be made
from the discussion above. They are: (A) $A$(Li)$_p$ is $\simeq$~0.10~dex below
that observed in HD~84937 and BD$+26^\circ3578$, using the $^6$Li observations
and depletion estimates to infer the underlying primordial $^7$Li value; (B) it
is the value measured in the most metal-poor star of the sample,
CD$-24^\circ17504$ at [Fe/H]~=~$-3.55$, whose metals reveal a
minimally-processed sample of early Galactic material; or (C) it is the
extrapolation of the metallicity trend to [Fe/H]~=~$-4$ where the most
metal-poor stars known are found and where the metallicity distribution of the
halo shows signs of truncation (Beers et al. 1998; Norris 1999).  The values
obtained are $A$(Li)$_p$~=~2.06~(A), 1.97~(B), and 1.98~(C).  That is, {\it we
infer that the primordial abundance is} $A$(Li)$_p$~$\simeq$~2.00, and that
future measurements of {\it stars with [Fe/H]~$<$~$-3.0$ will yield values of
$A$(Li) lower than the bulk of the present sample} (for which
$A$(Li)~$\simeq$~2.1), concordant with the trend shown in Fig.~8(a).  What
uncertainties should we attach to our estimate of the primordial value?  We
refer readers to the comprehensive discussion of errors by Thorburn (1994, \S5)
and to our previous works (Norris et al. 1994; Ryan et al. 1996a), and
summarise below the results most relevant to the present discussion.

We have previously noted that typical random errors in our estimation of
$A$(Li) are $\sigma_{\rm err} = 0.033$~dex.  Amongst systematic errors,
Thorburn gives the uncertainties in oscillator strengths as $\sigma \simeq
0.02$~dex.  A 0.5~dex error in lg~$g$ would produce $< 0.01$~dex error in
$A$(Li) at the turnoff. Reasonable uncertainties in microturbulence and the
damping coefficient are similarly unimportant due to the dominance of thermal
broadening in the core of this weak line of a species with such low atomic
mass.  Corrections for NLTE are $-$0.01 at 6100~K and $-$0.03~dex at 6300~K
(for [Fe/H]~=~$-2$ and lg~$g$~=~4; Carlson et al. 1994).  Far greater
systematic uncertainties arise due to the uncertainties in the zeropoint of the
effective temperature scale and the model structures. In \S5.3, we found it
necessary to make zeropoint adjustments to the various scales by as much as
165~K, which for the turnoff stars corresponds to an $A$(Li) change of
0.11~dex.  A similar difference arises in the abundances derived from the Bell
models compared with those from Kurucz's (1993) convective overshoot models,
the latter giving $A$(Li) higher by 0.08~dex at the turnoff (Ryan et al. 1996a,
\S3.3).  Bonifacio \& Molaro's (1998) study of the Li~6140~\AA\ line in
HD~140283 shows that abundances derived from the 6707~\AA\ resonance doublet
are not grossly in error.

Since $\sigma_{\rm obs} = 0.031$~dex and $\sigma_{\rm err} = 0.033$~dex, we
have established that there is no intrinsic spread about the Li plateau at the
metal-poor turnoff, to a level $\sigma_{\rm int} < 0.02$~dex.  It is clear that
the absolute uncertainties in the primordial abundance are dominated not by
random errors but by four systematic factors: (1) the zeropoint in the
metal-poor effective temperature scales, $\simeq 0.1$~dex; (2) uncertainties in
the metal-poor model atmosphere structures, $\simeq 0.1$~dex; (3) correction of
the observed level for the contamination of GCE $^6$Li and GCE $^7$Li; and (4)
correction for any destruction of pre-stellar Li.  Our three approaches (above)
to account for the GCE fraction gave results ranging over 0.09~dex.  That is,
sources (1), (2), and (3) each contributes $\simeq$~0.1~dex to the systematic
uncertainty in $A$(Li)$_p$~$\simeq$~2.00.

Until recently, source (4) was perhaps the most uncertain, since the degree of
depletion predicted by models depends very much on the input physics. The
simplest models predict essentially no destruction of Li ($<$0.05~dex) at the
metal-poor turnoff (Deliyannis et al. 1990), whereas rotationally-induced
mixing led Pinsonneault et al. (1998) to infer destruction by 0.2--0.4~dex.
However, the observations presented in this work set much tighter constraints
on the degree of rotationally-induced mixing than the data available to
Pinsonneault et al. could do, and on the basis of the very narrow scatter we
have measured, we conclude that depletion by the rotationally-induced mixing
mechanism is $<$~0.1~dex.  Although this limit is more severe than Pinsonneault
et al. were able to establish, it is consistent with Fields \& Olive's (1998)
limit of $<0.2$~dex depletion of $^7$Li, argued on the basis of light isotope
ratios.

Observations show that diffusion has not affected $A$(Li) (Ryan et al. 1996a),
but it is unclear how diffusion is inhibited. Vauclair \& Charbonnel (1995)
suggest that {\it small} stellar winds balance diffusive effects while avoiding
nuclear burning.  Although the simplest models present an incomplete picture
and fail to explain many behaviors (e.g. Deliyannis 1995), they may yet be
giving the correct result for the turnoff stars. Certainly the thinness of the
Li plateau argues against the models with rotationally-induced mixing, for
which a larger spread in $A$(Li) is predicted. Economy of hypothesis in this
situation suggests that systematic error source (4) is rather small.  However,
Vauclair (1999) challenges empirical inferences of this sort in the face of
current models in which depletion seems unavoidable.  Another possibility
requiring further study is discussed in \S7.8.

We finish this section by noting that the essentially zero scatter found for
the very metal-poor turnoff stars points strongly towards there having been a
primordial value for $^7$Li, and near-elimination of the concerns over its
depletion in these stars (but see Vauclair 1999 for an opposite view) suggests
that we are now closer to identifying that value with confidence.  Burbidge \&
Hoyle (1998) have considered that of the three factors: ({\it a}) stellar
processing, ({\it b}) Galactic production, and ({\it c}) Big Bang
nucleosynthesis, ({\it c}) is the one that has not operated.  The results of
the current study drive us to the contrary conclusion that ({\it a}) has not
operated significantly, ({\it b}) can be constrained jointly by the $^6$Li
abundance of these objects and the measured dependence of $A$(Li) on [Fe/H],
and that ({\it c}) is the most likely cause for the near-uniformity first
reported by Spite \& Spite (1982), supporting their conclusion that the
observed abundance was ``hardly altered'' from the primordial one.

\subsection{The Spread in Lithium Abundances in Globular Clusters}

A discussion of the spread of Li in field stars would not be complete without
reference to the observations of Li in subgiants in the globular cluster M92
which show a range of $A$(Li) (Deliyannis et al. 1995; Boesgaard et al. 1998).
Those authors considered whether various Li production mechanisms --- the
neutrino process in SN~II, Galactic cosmic ray $\alpha + \alpha$
nucleosynthesis, and $^7$Be transport in AGB stars --- could account for the
diversity, but in each case found requirements that violated other
observational constraints, such as expectations of enhanced [Mg/Fe] ratios, age
spreads within the cluster itself, and enhanced abundances of s-process
elements. They were driven  to prefer scenarios in which the range of $A$(Li)
reflected differential depletion from a higher abundance, rather than
differential enhancement from a lower level.

It is perhaps surprising that M92 reveals a spread in abundance of a factor of
2--3 for a small sample of stars, whereas in the field we find no spread
($\sigma_{\rm int} < 0.02$~dex) for 91\%\ of our sample.  The mean
metallicities of our samples are not greatly dissimilar, and the stellar masses
must be almost identical since our sample is right at the turnoff and the M92
sample is on the subgiant branch.  Moreover, the globular cluster sample should
have an even narrower age distribution than the field sample. Either some
feature of the globular cluster environment or the different post-main-sequence
evolution of the subgiants must be responsible for the differences, assuming
both data sets are reliable.

Can we reconcile Boesgaard et al's preference for a rotationally-induced
depletion mechanism in the globular clusters with the absence of a spread in
the halo? Possibly.  If environmental factors are responsible for the
difference, we may question whether the globular cluster members experience a
very different history of angular momentum evolution, giving rise to a larger
spread in $A$(Li).  Certainly the suggestion of different angular momentum
distributions between cluster and field star samples is not new.  Peterson,
Tarbell, \& Carney (1983) and Peterson (1983) first demonstrated that the
projected rotational velocities of horizontal branch stars in globular
clusters, having values of $v$sin$i$ up to ~30 km~s$^{-1}$, are significantly
higher than in their field counterparts, and speculation has long existed that
the ubiquitous chemical abundance anomalies in globular clusters (which in many
cases appear to have a bimodal signature) and which are absent among halo field
stars, are also driven by different angular momentum profiles (Norris 1981;
Suntzeff 1981).  While no satisfactory model currently exists to explain the
rich and somewhat bewildering literature on globular cluster abundance
variations (now known to involve C, N, O, Mg, Na, Al, Ba, and Eu (see Sneden et
al. 1997, and references therein\footnote{For simplicity we exclude from
discussion the even more complicated abundance patterns of the cluster $\omega$
Centauri (Norris \& Da Costa 1995).})), the signatures of abundance variations
have been found even at or near the main sequence turnoffs of some clusters
(eg. 47 Tuc (Briley et al. 1996), NGC 6752 (Suntzeff \& Smith 1991), and, most
importantly in this context, M92 (King et al. 1998)).  In M92, King et al.
report ranges in the abundances of Mg, Na, and Ba in the same stars for which
Li variations have been found, though they were unable to discern any
systematic correlation between the behavior of Li, on the one hand, and the
heavier elements, on the other.

While most efforts to understand the abundance anomalies have centered on the
angular momentum distribution within individual stars, under the supposition
that internal rotation might drive mixing, this provides an inadequate
explanation for the existence of variations at and below the main-sequence
turnoff (Da Costa \& Demarque 1982).  Alternatively one might speculate on
pre-main-sequence origins for the phenomenon, and interactions between crowded
protostellar disks have been proposed by Kraft (1998) as a possible mechanism
for generating different abundance patterns in cluster environments.  In this
context, then, is it possible that interactions between the disks in the dense
cluster environment enforce a diversity of evolutionary paths for the stars'
angular momenta, which then affect the Li profiles in these objects?  As most
of the Li depletion and dispersion in the rotationally-induced turbulent models
occurs during the first $<$ 0.3 Gyr (Pinsonneault et al. 1992, Fig. 7), it is
possible that the crowded cluster environments are affected by interactions
between protostellar disks at just this crucial, early phase, producing
different initial conditions to those found in lower density star clusters
which ultimately dissolved to form the field population.  If the environmental
conditions have given rise to different Li-processing histories and generated
different $A$(Li) spreads, then we must re-ask whether the thinness of the
field Li plateau signifies a lack of depletion or merely depletion under
conditions that were similar from one field star to another.

Alternatively, the $A$(Li) spread in M92 may be due to some other unidentified
cause, which may possibly also explain the high abundance in the field star
BD$+23^\circ3912$ (King, Deliyannis, \& Boesgaard 1996).  In the field
population, such enigmatic stars appear to be even less common than the
ultra-Li-depleted stars, so it may be appropriate to regard them (or ``it'') as
rare pathological cases not requiring us to lose sight of the ``health'' of the
majority of Li plateau stars. Purists may argue, with some merit, that the
Population II lithium origin cannot be determined with certainty until all such
exceptions are understood.  The observations in M92 raise the interesting
possibility that globular cluster stars may exhibit quite different Li
processing histories than the field stars. We stand to learn more not only
about Li but also about the differences in globular cluster and low density
cluster environments from more detailed study, at higher S/N, of additional
stars in this and other globular clusters.

\section{Concluding Remarks}

The vast majority (91\%) of our very metal-poor, main-sequence turnoff, field
sample is consistent with an observed scatter of only $$\sigma_{\rm
obs}~=~0.031~{\rm dex}$$ about a mean $A$(Li)~=~2.11~dex.  G186-26, being
ultra-Li-depleted, was rejected (ab initio) from the analysis; it is a reminder
that some stars deplete their Li by 1~dex or more.  BD+9$^\circ$2190 was
rejected by the outlier-detection algorithm from the ``best'' sample on account
of an anomalous abundance compared to the other stars. Even so, the larger
formal errors associated with this star make it unclear whether it is genuinely
depleted, or merely an inferior observation.  Its inclusion in the ``best''
sample would have inflated the observed scatter to only 0.037~dex, compared to
the expected errors $\sigma_{\rm err}~=~0.033$~dex, so irrespective of its
status we conclude that the intrinsic scatter of $A$(Li) for the metal-poor
turnoff is $$\sigma_{\rm int} < 0.02~{\rm dex}.$$

We have again found a strong dependence of $A$(Li) on metallicity, $${\rm
d}A{\rm(Li)}/{\rm d [Fe/H]}~=~0.118 \pm 0.023~{\rm dex~per~dex},$$ which is
concordant with theoretical GCE models and with observed $^6$Li levels.

Four systematic uncertainties are discussed. Three involve the adopted
temperature scale, the atmospheric models, and interpretation of the GCE
contamination revealed by $^6$Li and the metallicity trend.  These systematic
uncertainties are $\simeq~0.10$~dex in each instance.  In this study, the
effective-temperature zeropoint was set by Magain's (1987) and Bell \& Oke's
(1986) $b-y$ calibrations of metal-poor stars, and the model atmospheres are
from Bell (1983), which do not possess the convective overshoot used in
Kurucz's (1993) models.  The fourth systematic uncertainty surrounds possible
stellar depletion of the pre-stellar Li.  The inferred intrinsic scatter, if
any, must be essentially zero, $\sigma_{\rm int} < 0.02$.  This is much less
than the range expected for the rotationally-induced turbulent mixing mechanism
of Pinsonneault et al. (1998), and we conclude that depletion by that mechanism
must be $< 0.1$~dex.  If essentially no surface Li has been destroyed in these
very metal-poor turnoff stars, then the only substantial correction required to
the mean abundance is for GCE, leading to a primordial abundance {\it lower}
than the plateau mean. We infer $A$(Li)$_p$~$\simeq~2.00$~dex. The three
surviving, potential systematic uncertainties listed at the beginning of this
paragraph are $\simeq$~0.10~dex each.

The difference between our field star observations and the M92 data of
Boesgaard et al. (1998) suggests that real field-to-cluster differences in Li
evolution may have occurred. These may indicate different angular momentum
evolutionary histories, possibly associated with interactions between
protostellar disks in the dense globular cluster environments.  Further
accurate study of Li in globular clusters will be required.

\acknowledgements

The authors gratefully acknowledge discussions with Dr C. P. Deliyannis and Dr
J. A. Thorburn on an earlier, similar, proposal that was not supported by the
telescope time assignment committees. They are grateful to Dr W. J. Schuster
for supplying new Str\"omgren photometry ahead of publication, and to Dr A.
Pedrosa for obtaining the WHT service observation.  They also record their
thanks to the Director and staff of the Anglo-Australian Observatory and the
Australian Time Assignment Committee for the provision of facilities during
this investigation.  T.C.B. acknowledges partial support for this work from
grants AST 90-1376, AST 92-22326, INT 94-17547, and AST 95-29454 awarded by the
National Science Foundation.

\vfill
\eject

{\bf Figure captions}
 
Fig. 1: 
Spectra in region of the Li 6707~\AA\ line, offset by multiples of 0.1 continuum
units.
Multiple epochs have been co-added for this illustration, and the continuum
location has been indicated with a dotted line, but actual 
measurements of equivalent widths were made for each epoch separately, to
check repeatability. See text for details.

Fig. 2: 
Comparison of equivalent width measurements for stars in common to our work and
the accurate isotope ratio work of Smith et al. (1998). Our data are in 
agreement with theirs within $\pm 1.8\sigma$ at worst, and considerably better
in many cases. The {\it dotted line} is the 1:1 locus.

Fig. 3: 
Dereddened Str\"omgren $c_1^0$ vs $(b-y)_0$ diagram showing our
turnoff sample ({\it solid symbols}) against the general halo sample with
[Fe/H] $< -1.0$ of 
Schuster, Parrao, \& Contreras Martinez (1993) ({\it crosses}).

Fig. 4: 
Dereddened indices available to measure effective temperature, as a function of
$(b-y)_0$.
(a) The solid line coinciding with the data is from the theoretical colors
of Bell \& Oke (1986) for [Fe/H]~=~$-$2 and lg~$g$~=~4.0.
The solid line sitting away from the data is the
transformation of Magain (1987) at [Fe/H]~=~$-2.8$. 
(b) and (c) {\it Solid lines}: Bell \& Oke theoretical colors.
(d) and (e) {\it Solid lines}: least squares fits of $(b-y)_0$ to the index, 
used to predict $(b-y)_0$ from the measured indices.

Fig. 5:
Metallicity sensitivity of color-effective temperature calibrations for
metal-poor turnoff stars:
{\it solid curve}: synthetic colors of Kurucz (1993);
{\it dashed curve} ((a) and (d) only): empirical fit to IRFM temperatures by Magain (1987); 
{\it dotted curve}: empirical fit to IRFM temperatures by Alonso et al. (1996a).
Pairs of values are traced for each color. 
In the Kurucz and Magain calibrations, the sensitivity to metallicity decreases 
as expected as [Fe/H] falls from $-1$, whereas the Alonso et al. calibrations
go through a minimum before increasing non-physically towards yet lower 
metallicity.
(a) B--V = 0.35 and 0.40. 
The {\it solid bar} shows the metallicity range of our sample.
(b) (V--R)$_{\rm C}$ = 0.26 and 0.28 (which were transformed to Johnson 
colors for Alonso et al's calibration).
(c) (R--I)$_{\rm C}$ = 0.29 and 0.31 (which were transformed to Johnson 
colors for Alonso et al's calibration). 
(d) $b-y$ = 0.29 and 0.32.  
Alonso et al's calibration appears highly non-physical over the range
[Fe/H]~$<$~$-1$. Their $b-y$~=~0.29 curve is shown for two values of 
$c_1$~=~0.32 and 0.38; the lower curve is for $c_1$~=~0.30. 
(e) $\beta$~=~2.60 and 2.62 for Alonso et al. calibration, and
$\beta$~=~2.65 and 2.66 for Kurucz calibration.

Fig. 6:
(a) Equivalent widths vs effective temperature.
(b) Equivalent widths on lg scale, which is linear in $A$(Li).
Solid line is for $A({\rm Li})~=~2.11$.
(c) Spread in $A$(Li) about the 2.11~dex locus. Dashed lines are at
$\pm$0.072~dex (2 s.d.) from the mean of the majority.
(d) Histogram (upper) and stripe plot (lower) of $A$(Li) spread. 
The sample is seen to consist of a well
defined bell curve to which the majority of the data conform, plus two stars
lower in $A$(Li) by $\sim$~0.14~dex. See text for discussion.

Fig. 7:
Spread in standardised residuals 
($Z_i = (A_i({\rm Li}) - 2.11)/\sigma_{A({\rm Li}),i}$) about the 2.11~dex 
locus. 
See text for discussion.

Fig. 8:
(a-d) ``Best Fit'' sample (G186-26 rejected a priori).
(a) Dependence of $A({\rm Li})$ on [Fe/H]. 
{\it Dotted line}: Univariate OLS fit for all stars;
{\it solid line}: Univariate RWLS fit in which BD+9$^\circ$2190 was 
rejected by the analysis as an outlier. This is our ``best fit'' regression.
(b) $A$(Li) residuals from best fit.
(c) Distribution of residuals shown as both a conventional histogram and a
stripe plot.
(d) ``Normal probability plot'' confirming that the residuals are 
distributed normally (see text). {\it Dotted line}: OLS fit to guide the eye,
to highlight linearity.
(e-h) ``Culled Fit'' sample from which BD+9$^\circ$2190, CD$-24^\circ$17504 and
CD$-71^\circ$1234 have been excluded, illustrating that even a culled sample
yields a significant metallicity dependence. 
(e) Dependence of $A({\rm Li})$ on [Fe/H].
{\it Solid line}: Univariate RWLS fit in which BD+9$^\circ$2190 was
excluded a priori, and subsequently CD$-24^\circ$17504 and
CD$-71^\circ$1234 were rejected by the analysis as outliers.
(f) As for (b).
(g) As for (c).
(h) As for (d).

Fig. 9:
(a-d) Lithium abundances derived for two sets of effective temperature values,
and plotted against two sets of [Fe/H] values; 
{\it solid line} = OLS fit.
(e-f) 
Metallicities derived from 1\AA -resolution spectra compare well with the
high-resolution values adopted in this study, but unfavourably with several
values adopted by Bonifacio \& Molaro (1997). 
{\it Dotted line} = 1:1 locus for different [Fe/H] scales.
\newline
Due to the large scatter 
introduced by the IRFM temperatures and the disagreement between the 
[Fe/H] values from 1\AA-resolution spectra and the values adopted by
Bonifacio \& Molaro, we argue that panel (a) is the most reliable presentation
of the data. See text for discussion.

Fig. 10:
(a) Depletion curves from Pinsonneault et al. (1998) ``s0'' model,
broadened for formal errors of 0.033~dex. 
{\it Solid curve}: median depletion;
{\it dashed curves}: boundaries enclosing $\pm$47.5\%\ of the population.
The observational data have been superimposed for an 
assumed initial $A$(Li)$_i$~=~2.23.
(b) {\it Shaded histogram}: ``s0'' model renormalized to our sample size,
assuming $A$(Li)$_i$~=~2.23;
{\it heavy histogram}: observations uncorrected for embedded metallicity 
dependence.
(c) {\it Shaded histogram}: ``s0'' model as for (b);
{\it heavy histogram}: observations offset to [Fe/H]~=~$-2.8$ to compensate
for the embedded metallicity dependence. The model predicts a Li-depleted tail 
comprising 17\%\ of the sample, but it is not populated by the observations.

%
%
%
%
%
%
%
%
%

\end{document}